\documentclass[aps,english,showpacs,twocolumn,pra,superscriptaddress]{revtex4-1}
\usepackage{amsfonts}
\usepackage{amssymb}
\usepackage[colorlinks=true,citecolor=blue,linkcolor=magenta]{hyperref}
\usepackage{amsmath}
\usepackage{graphicx}
\usepackage{epsfig}
\usepackage{color}
\usepackage{float}
\usepackage{ulem}

\begin{document}

\title{Expressibility of linear combination of ansatz circuits}
\author{Peng Wang}
\affiliation{Department of Mathematics and Physics, North China Electric Power University, 102206 Beijing, People’s Republic of China}
\author{Ruyu Yang}
\email{yangruyu96@gmail.com}
\affiliation{Graduate School of China Academy of Engineering Physics, Beijing 100193, China}

\begin{abstract}
Variational Quantum Eigensolver is considered promising for medium-scale
noisy quantum computers. Expressibility is an important metric for measuring
the capability of a variational quantum Ansatz circuit. A commonly used
method to increase expressibility is to increase the circuit depth. However,
increasing the circuit depth also introduces more noise. We propose to use a
linear combination of ansatzes to improve the expressibility of variational
circuits, thus avoiding the increase of circuit depth. Concurrently, we introduce a novel measurement strategy that circumvents the necessity for the Hadamard test, thereby significantly diminishing the reliance on two-qubit gates, which are presently the predominant contributors to quantum noise. We also provide a corresponding gradient calculation method, which
makes it convenient to update the parameters. Compared with the method of
increasing the circuit depth, our method of improving expressibility is more
practical. Numerical simulations demonstrate the effectiveness of our method.
\end{abstract}

\maketitle

\affiliation{Department of Mathematics and Physics, North China Electric Power
University, Beijing 102206, China}

\affiliation{Graduate School of China Academy of Engineering Physics,
Beijing 100193, China}

\section{INTRODUCTION}

The physical realization of quantum computing has undergone rapid
development in recent years, and scientists are now able to create quantum
computers with over one hundred qubits in the laboratory ~\cite%
{ebadi2021quantum}. Due to limitations in the number of qubits and the precision of quantum gates, we are currently in the Noisy Intermediate-Scale
Quantum (NISQ) era~\cite{preskill2018quantum,lau2022nisq,abughanem2023nisq},
and there is still a way to go before fault-tolerant quantum computing can
be achieved. An important example of an algorithm suitable for NISQ quantum
computers is Variational Quantum Eigensolver (VQE) ~\cite%
{peruzzo2014variational,wecker2015progress,mcclean2016theory,tilly2022variational,cerezo2021variational,fedorov2022vqe}%
. VQE itself is a broad and active research field. From an application
perspective, VQE can be used to solve problems in quantum chemistry~\cite%
{deglmann2015application,cao2019quantum,outeiral2021prospects,mcardle2020quantum,google2020hartree}
, condensed matter and materials physics~\cite%
{dallaire2019low,bravo2020scaling,xu2020test,PhysRevB.102.075104}, and
nuclear physics~\cite%
{PhysRevB.102.075104,PhysRevA.100.012320,PhysRevD.101.074038}. VQE itself is
not a deterministic algorithm. An important factor affecting the performance
of VQE is the design of the ansatz circuit. In order to make VQE achieve
better performance, various ansatz have been designed~\cite%
{2017Hardware,2019Noise,anand2022quantum,PRXQuantum.1.020319,grimsley2019adaptive,gard2020efficient,grimsley2019trotterized}%
. In Ref~\cite{sim2019expressibility}, the authors propose the concept of
expressibility of parameterized quantum circuits. In addition to this
definition method, there are other definitions of expressibility~\cite%
{funcke2021dimensional,du2022efficient}. Intuitively, expressibility
measures the ability of a parameterized quantum circuit to generate quantum
states. Generally speaking, when little is known about the target state, the
higher expressibility the ansatz circuit is, the more likely it is that it
covers the potential target state. Only when the target state is covered it
can be possible to solve the corresponding problem through VQE. A natural
way to improve the {expressibility} is to increase the depth of the ansatz
circuit~\cite{sim2019expressibility,du2022efficient}. However, in the NISQ
era, increasing quantum circuit depth will also increase noise and reduce
circuit fidelity. This significantly increases the resources required to
mitigate these noises~\cite%
{PhysRevLett.119.180509,li2017efficient,endo2018practical}.

To overcome this difficulty, we propose to improve expressibility through
linear combinations of different ansatzes. Intuitively, using the linear
combination of ansatzes (LCA) as a new ansatz can greatly increase the
number of trainable parameters without increasing the depth of the circuit.
Correspondingly, more quantum states can be reached by changing the
parameters {than single ansatz (SA). Our numerical results provide evidence for that and exhibit
LCA has higher expressibility than SA. For a fixed system size, increasing
the ansatzes number in linear combination can rapidly improve the expressibility without increasing the circuit depth, until the ansatzes number reaches the threshold that is linear
to the system size.}

In the past, the implementation of unitary linear combinations often relied
on the Hadamard test. However, the Hadamard test relies on the
implementation of the Control-$U$ gate (in the LCA algorithm, $U$ is an
ansatz circuit). In the circuit that implements Control-$U$ gates, a
single-qubit gate in $U$ becomes a two-qubit gate, and a CNOT gate becomes a
Toffoli gate. This means that the Hadamard Test circuit requires a large
number of two-qubit gates, which are the main source of noise in quantum
circuits. To further reduce the impact of noise, we propose a new
sub-algorithm, named phase consistent measurement (PCM) for LCA to replace
the Hadamard test. PCM avoids the use of Control-$U$ gates, thereby reducing
the use of two-qubit gates. PCM relies on the observation that the phase
difference between the final states corresponding to different ansatz does
not affect the results of VQE. Because we never derive the phase difference
between the final states when calculating the average value of the
Hamiltonian, the commonly used parameter shift rules cannot be applied
directly~\cite%
{kyriienko2021generalized,guerreschi2017practical,mitarai2018quantum,PhysRevA.99.032331,bergholm2018pennylane}%
. Based on our specific question, we propose a corresponding parameter
update scheme that works with common ansatz to reduce the number of
measurements required to update parameters.

In Sec.~\ref{PRELIMINARIES}, we revisit the core concept employed in this article. Subsequently, in Sec.~\ref{Expressibility of LCA}, we numerically investigate the Expressibility of LCA. Moving on, Sec.~\ref{implement} demonstrates the implementation of LCA without ancilla qubits, highlighting the reduction in the number of CNOT gates required. Lastly, we provide a comprehensive summary of this work in Section~\ref{Summary}.
\label{Introduction}

\section{PRELIMINARIES}

\label{PRELIMINARIES}

\subsection{The linear combination of ansatz}

\label{Expressibility}
\begin{figure*}[t]
\centering
\includegraphics[bb=0 0 2525 1850, width=17 cm, clip]{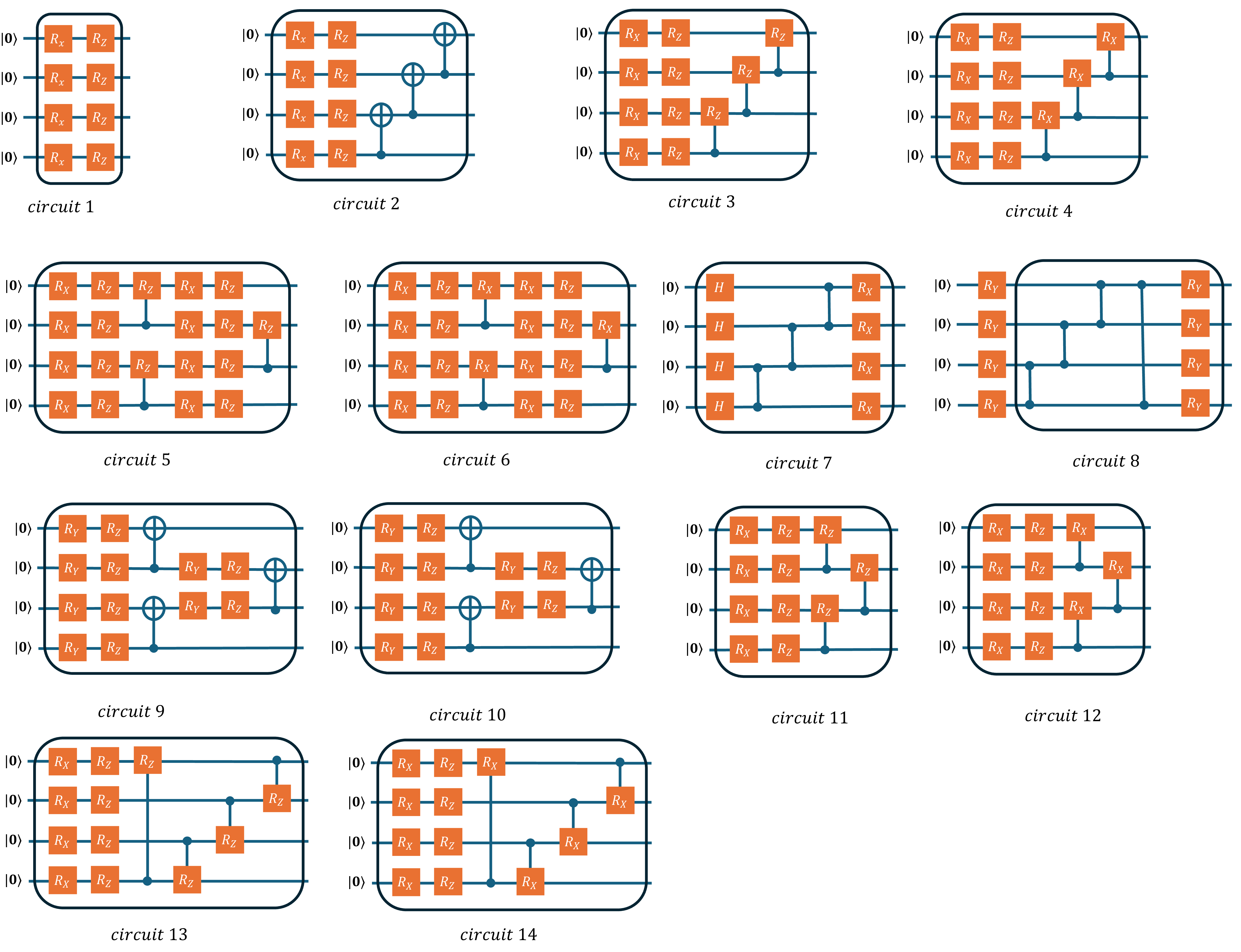}
\caption{{Schematic diagram of $14$ ansatzes of which the linear combination
is studied in this work. The box represents a layer of the circuit. These
circuits can be generalized to multi-qubit cases.}}
\label{14circuits}
\end{figure*}
\label{Combined-ansatz variational quantum eigensolver} Ansatz is the key
part of VQA. It is a parameterized quantum circuit $U(\boldsymbol{\theta })$
generating a trial state $\left\vert \psi _{\boldsymbol{\theta }%
}\right\rangle =U(\boldsymbol{\theta })\left\vert \phi _{0}\right\rangle $,
where $\boldsymbol{\theta }$ is a parameter vector $\boldsymbol{\theta =}$\{$%
\theta _{1},\theta _{2},...$\}, $\left\vert \phi _{0}\right\rangle $ is a
reference state. VQA minimizes the cost function $C=\left\langle \psi _{%
\boldsymbol{\theta }}\right\vert A\left\vert \psi _{\boldsymbol{\theta }%
}\right\rangle $ by training the parameter vector $\boldsymbol{\theta }$
according to the formula%
\begin{equation}
\theta _{k}\rightarrow \theta _{k}-\alpha \partial _{k}C,
\end{equation}%
where $\partial _{k}C\equiv \partial C/\partial \theta _{k}$ and $\alpha$ is the learning rate, until the cost
function approaches the exact result with acceptable accuracy.

In this article, we introduce the combined ansatz, i.e. the linear
combination of several ansatz circuits. Given a set of ansatz circuit \{$%
U^{i}(\boldsymbol{\theta }^{i}),i=1,\cdots ,M$\}, the combined ansatz is
\begin{equation}
\widetilde{U}(\boldsymbol{\tilde{\theta}})=\frac{1}{\Omega }%
\sum_{i=0}^{M-1}c_{i}U^{i}(\boldsymbol{\theta }^{i}),  \label{LCU}
\end{equation}%
where $\boldsymbol{c}=\{c_{1},c_{2},\mathbf{...,}c_{M}\mathbf{\}}$ is the
vector of complex coefficients, $\boldsymbol{\tilde{\theta}}=\{{\boldsymbol{
\theta }}^{1},{\boldsymbol{\theta }}^{2},\cdots ,{\boldsymbol{\theta }}
^{M},c\}$ is the set of parameter vectors. $\Omega $ is the normalization
coefficient and determined through $\left\langle \phi _{0}\right\vert (
\widetilde{U}(\boldsymbol{ \tilde{\theta}}))^{\dagger }\widetilde{U}(
\boldsymbol{\tilde{\theta}})\left\vert \phi_{0}\right\rangle =1.$ Note that
Re($c_{i}$)$,$ Im($c_{i}$) and $\boldsymbol{\theta }^{i}$ ($i=1,2,...$) are
trainable. {The Hardware efficient ansatz (HEA) is a type of ansatzes that
are easily implementable and not tailored to specific problems. In this
work, we adopt 14 HEAs described in Ref.~\cite{sim2019expressibility} as $%
U^{i}$ to construct the linear combination of ansatzes. The schematic
diagram of the 14 HEAs with system size $Q=4$ is shown in Fig.~\ref%
{14circuits}.}

\subsection{Expressibility}

\label{Expressibility}
\begin{figure}[t]
\centering
\includegraphics[width = 0.8\linewidth]{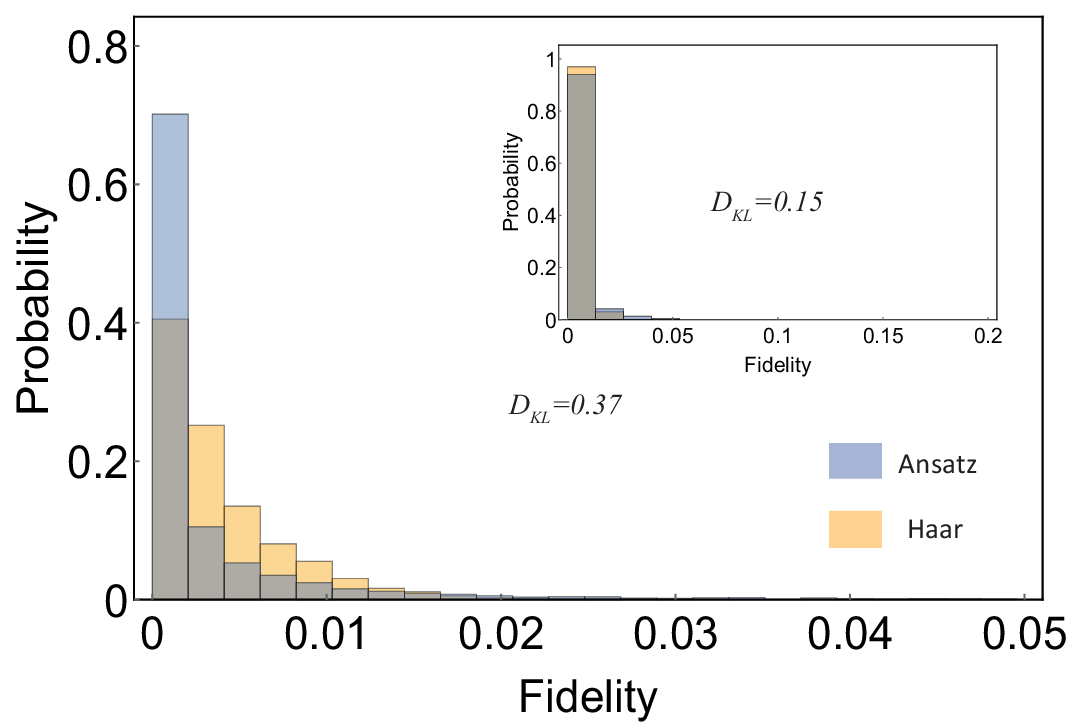}
\caption{Histograms of estimated fidelities are shown, overlaid with
fidelities of the Haar-distributed ensemble, with the computed KL
divergences. }
\label{histogram}
\end{figure}

The concept of expressibility serves as a key metric for evaluating the
efficacy of ansatz circuits. An ansatz circuit is deemed "good" for a given
problem if the generated ensemble of trial states harbors the true solution.
However, in many practical scenarios, access to the solution of the problem
might be limited. As an alternative, we use {expressibility} to evaluate the
quality of parameter circuits. Expressibility can be defined as the
capability to generate pure states, encompassing the ability to
comprehensively and uniformly explore the entire unitary space. In this
paper, we use Kullback-Leibler (KL) divergence to quantify expressibility~\cite{sim2019expressibility},{\
\begin{equation}
\text{Expr}=D_{\mathrm{KL}}(P_{\text{est}}(F)||P_{\text{Haar}}(F)),
\label{Expr}
\end{equation}
which measures the distance between two discrete probability distributions
of fidelity. $P_{ \text{est}}(F)$ is the estimated probability distribution
of the fidelity $F=\left\vert \left\langle \psi _{\phi }\right. \left\vert
\psi _{\theta }\right\rangle \right\vert ^{2},$ where $\left\vert \psi
_{\theta }\right\rangle $ and $\left\vert \psi _{\phi }\right\rangle $ are
obtained by independently sampling a pair of parameter vectors in ansatz
circuits. $P_{\text{Haar}}(F)$ is the probability distribution of the
fidelity $F$\ for the Haar-distributed random states, which can be obtained
by the analytical form $P_{\text{Haar}}(F)=(N-1)(1-F)^{N-2}$, where $N=2^{Q}$
is the dimension of the Hilbert space. The discretization implies splitting
the probability distributions into $n_{\mathrm{bin}}$ parts, then the value
of $D_{\mathrm{KL}}$ can be numerically calculated according to the
definition $D_{\mathrm{KL}}(P||P^{\prime })=\sum_{i}^{{n_{\mathrm{bin}}}}{P}%
_{i}\ln ({P}_{i}/P_{i}^{\prime }).$ $D_{\mathrm{KL}}${\ can be visualized
using a histogram with bin number of }$n_{\mathrm{bin}}$. Fig.~\ref%
{histogram} exhibits the histogram where $P_{\text{est}}$ are generated by
ansatz circuit $10$ with qubit number $Q=12$. }

The value of $D_{\mathrm{KL}}$ in Eq.~\eqref{Expr} depends on the bin number
of {$n_{\mathrm{bin}}$}. One can use a fixed bin number such as dividing $%
[0,1]$ into {$n_{\mathrm{bin}}$} parts and assign each part a value
according to the probability distribution~\cite{sim2019expressibility}.
However, in multi-qubit systems, the estimated probability distribution is
nonzero only in a narrow region, and the zero region does not contribute to $%
D_{\mathrm{KL}}$. Therefore, we use an unfixed bin number dividing the
nonzero region into {$n_{\mathrm{bin}}$} parts. {Fig.~\ref{histogram}} and
the inset imply that the histogram or $D_{\mathrm{KL}}$ with an unfixed bin
number more clearly distinguishes the difference between two probability
distributions than that with a fixed bin number when the bin size is larger
than the width of the nonzero region.

\begin{figure}[t]
\centering
\includegraphics[width=0.8\linewidth]{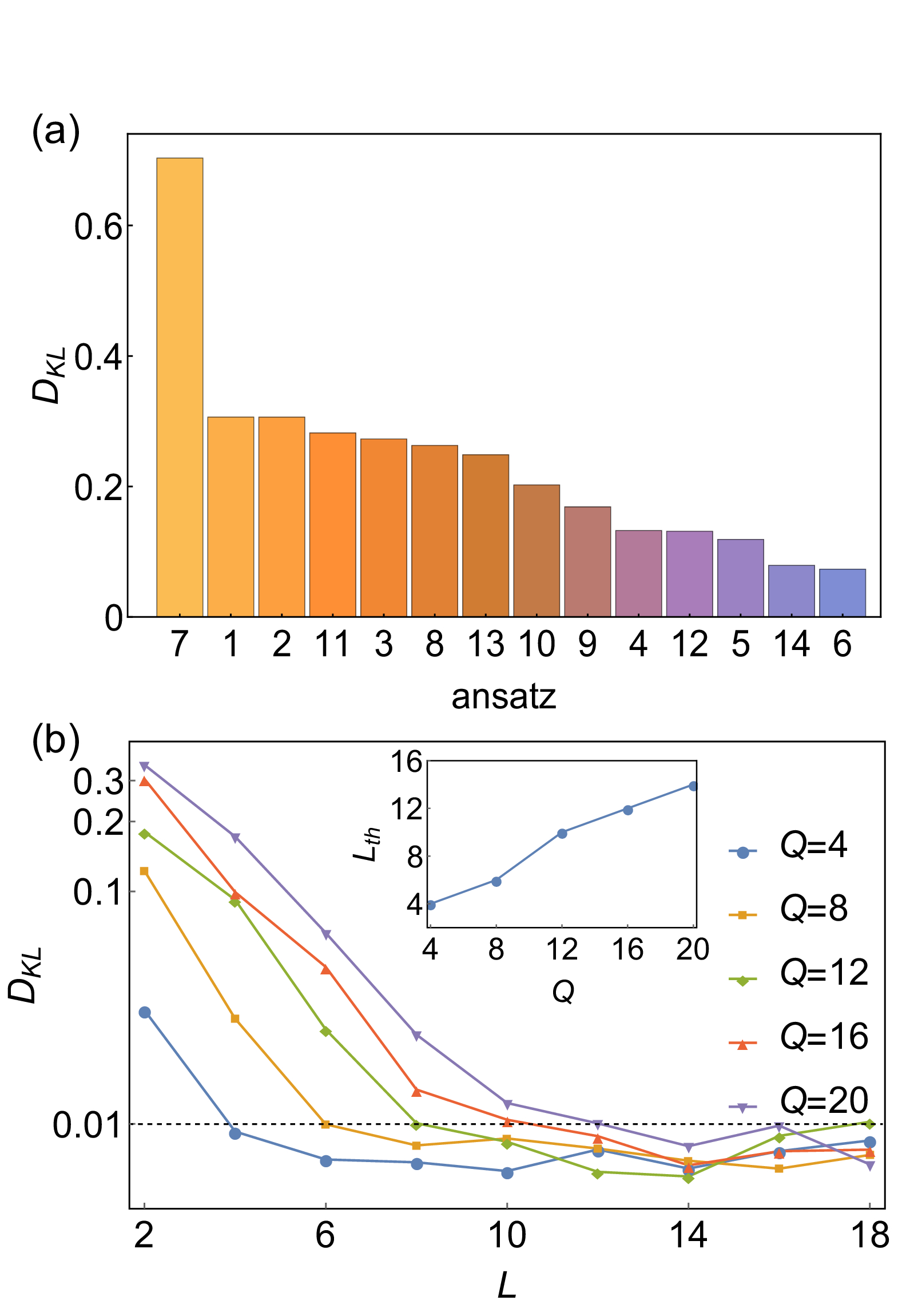}
\caption{{The numerical results of KL divergence (i.e., $D_{\mathrm{KL}}$)
with the unfixed bin number. (a) $D_{\mathrm{KL}}$ for single ansatz. The
x-axis represents the circuit ID. (b) $D_{\mathrm{KL}}$ as a function of the
circuit layer $L$ for qubit numbers $N=4,8,12,16$ and $20$. The black dotted
line serves as a reference line for the threshold of high expressibility.
The inset suggests that the threshold value is linear in the qubit number.}}
\label{DKLVaryingwithQandL}
\end{figure}

{The common method to enhance the expressibility is increasing the depth of
the ansatz circuit, such as repeating ansatz $U(\mathbf{\theta })$ of $L$
times. Taking circuit $10$ as an example, we shows the numerical results of $%
D_{\mathrm{KL}}$ varying with the repetitions number $L$ and system size $Q$
in Fig.~\ref{DKLVaryingwithQandL} b.} For a fixed system size, the
expressibility of $[U(\mathbf{\theta })]^{L}$ tends to stabilize after
reaching a threshold value of $L$. The inset suggests that the threshold $L_{%
\mathrm{th}}=aQ+b$ is linearly dependent on the system size, where $Q$ is
the qubit number, and $a$ and $b$ are constant. However, increasing the
number of repetitions $L$ will exponentially increase the noise, making the
system unusable due to excessive noise. In the following text, we will show
that LCA can improve expressibility with $L=1$ for any number of qubits.

\section{Expressibility of LCA}

\label{Expressibility of LCA}In this section, we extend the expressibility
defined in Eq.~\eqref{Expr} for SA to a linear combination of {\color{blue}LC%
$M$A}. The combined trial state, used for sampling the estimated
distribution is obtained by applying Eq.~\eqref{LCU} to the reference state:
\begin{equation}
\left\vert \psi _{\mathbf{\tilde{\boldsymbol{\theta }}}}\right\rangle =\frac{%
1}{\Omega }\sum_{i}^{M}c_{i}U^{i}(\mathbf{\boldsymbol{\theta }}%
^{i})\left\vert \phi _{0}\right\rangle .  \label{LC}
\end{equation}%
Here, $U^{i}(\mathbf{\theta }^{i})$ represents a hardware efficient ansatz.
We employ $14$ of $19$ HEAs studied in Ref.~\cite{sim2019expressibility}. {These
ansatzes in Ref.~\cite{sim2019expressibility} with high expressibility,
such as circuit $5,6,13,14,$ and unchanged expressibility as increasing
repeating layers, such as circuit $15$, are ignored. The expressibilities of
the chosen $14$ ansatzes are shown in Fig. \ref{DKLVaryingwithQandL}(a),
calculated with unfixed bin number. }For the sake of convenience, the
ansatzes involved in linear combinations in Eq. ({\ref{LC}}) are denoted by
the ansatz set $\bar{M}=\{i,j,k...\}$, where $i$ represents the $i$th HEA
and $M$ is the size of ansatz set. In the following subsections, we exhibit
the numerical simulation to compare the expressibility of SA and LC$M$A
through the calculation of $D_{KL}$ and ground state energy estimation.in
Fig. {\ref{DKLVaryingwithQandL} (b)}

\subsection{Demonstration with single-qubit case}

\label{Demonstration in single-qubit case}

In this subsection, we first present a single-qubit case to demonstrate that
the expressibility of LC$M$A is superior to that of SA.

\begin{figure}[t]
\centering
\includegraphics[width = 0.65\linewidth]{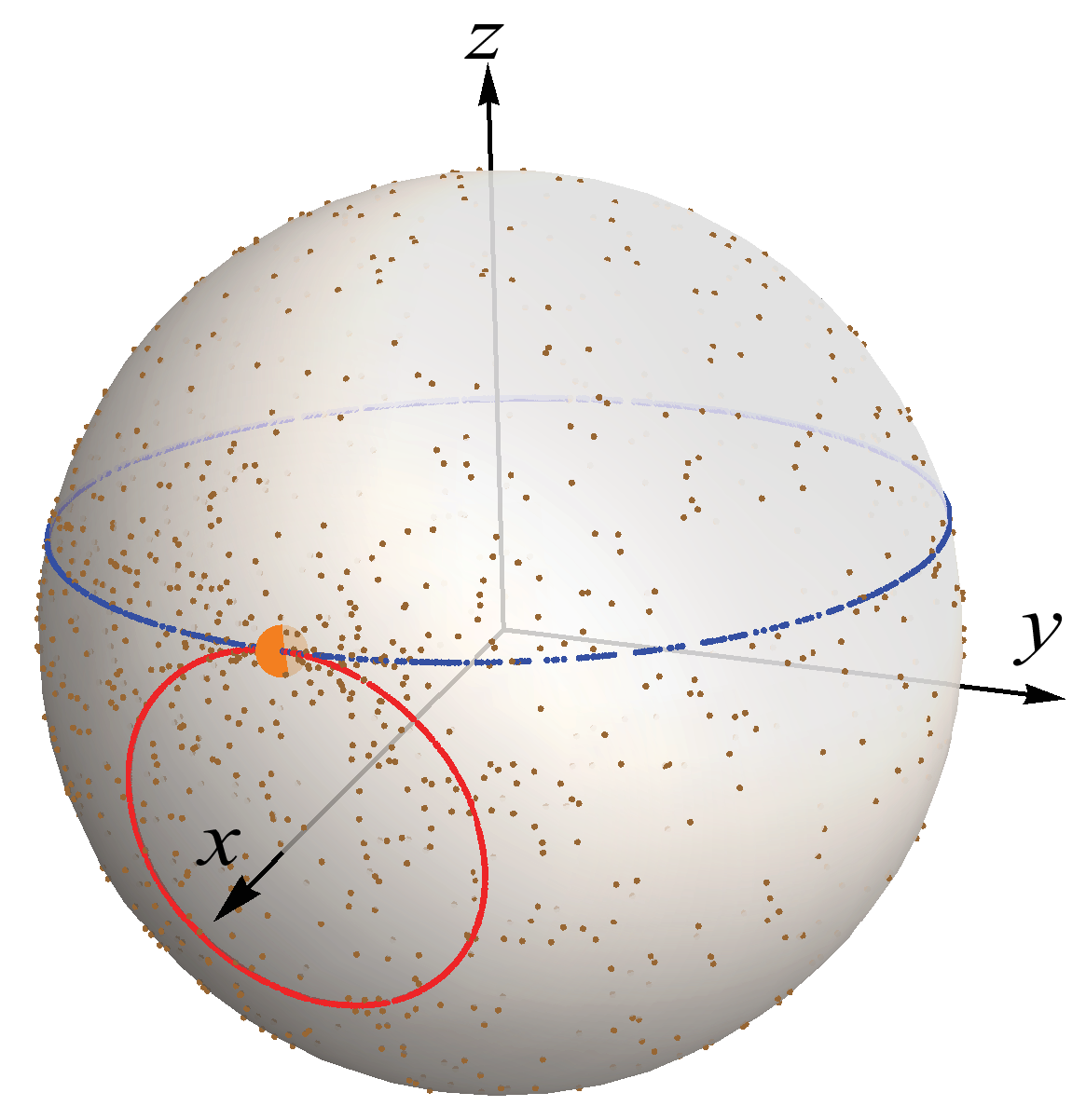}
\caption{$2000$ parameterized states were plotted on the Bloch sphere using
QuESTLink~\protect\cite{johansson2012qutip}. The red and blue data
respectively represent the parameterized states $R_{x}(\protect\theta )%
\protect\phi (0)$ and $R_{z}(\protect\theta ^{\prime })\protect\phi (0)$,
where $\protect\phi (0)=(\protect\sqrt{2}\left\vert 0\right\rangle
+\left\vert 1\right\rangle )/\protect\sqrt{3}.$ The brown data are the
parameterized states $[c_{1}R_{x}(\protect\theta )+c_{2}R_{z}(\protect\theta %
^{\prime })]/\Omega \protect\phi (0)$ where $\Omega $ is a normalization
coefficient.}
\label{bloch}
\end{figure}

Consider the scenario where $M=2$ with $U^{1}(\theta )=R_{x}(\theta )$ and $%
U^{2}(\phi )=R_{z}(\phi )$. Let the reference state be $\left\vert \phi
_{0}\right\rangle =(\sqrt{2}\left\vert 0\right\rangle +\left\vert
1\right\rangle )/\sqrt{3}$, depicted as the orange point in Fig.~\ref{bloch}%
. $1000$ sample pairs of parameterized states \{$U^{1}(\theta \mathbf{)}
\left\vert \phi _{0}\right\rangle \mathbf{,}U^{1}(\theta ^{\prime }\mathbf{)}
\left\vert \phi _{0}\right\rangle $\} and \{$U^{2}(\phi )\left\vert \phi
_{0}\right\rangle \mathbf{,}U^{2}(\phi ^{\prime })\left\vert \phi
_{0}\right\rangle $\} drawn in Fig.~\ref{bloch}. Notably, these
parameterized states form distinct circles (red for $U^{1}$ and blue for $%
U^{2}$) on the Bloch sphere, indicating a one-dimensional subspace
exploration. We calculate numerically the distribution of $1000$ fidelities $%
\left\vert \left\langle \psi _{\mathbf{\theta }^{\prime }}\right. \left\vert
\psi _{\mathbf{\theta }}\right\rangle \right\vert ^{2}$ and get $D_{KL}=2.18$
(for $U^{1}$) and $0.34$ (for $U^{2}$). As for the combined parameterized
states, we sample $1000$ pairs of parameter vectors \{$\mathbf{\theta }%
\mathbf{,\theta }^{\prime }$\} and obtain the corresponding $1000$ pairs of
parameterized states \{$\left\vert \psi (\mathbf{\theta })\right\rangle
\mathbf{,}\left\vert \psi (\mathbf{\theta }^{\prime })\right\rangle $\},
where $\mathbf{\theta }=\{\theta _{1},\theta _{2}\},\left\vert \psi (\mathbf{%
\theta })\right\rangle =1/\Omega (U^{1}(\theta _{1})+U^{2}(\theta
_{2}))\left\vert \phi _{0}\right\rangle $ and $1/\Omega $ is a normalization
coefficient. The combined parameterized states fill the Hilbert space, which
indicates that LCA can explore more space than SA. This coincides with the
numerical result $D_{KL}=0.047$ indicating the LCA has a higher
expressibility than SA.

\subsection{The improvement of the expressibility for two ansatzs}

\label{The improvement of the expressibility for two ansatzs}
\begin{figure}[t]
%\centering
\includegraphics[bb=0 0 693 690, width=8.5 cm,clip]{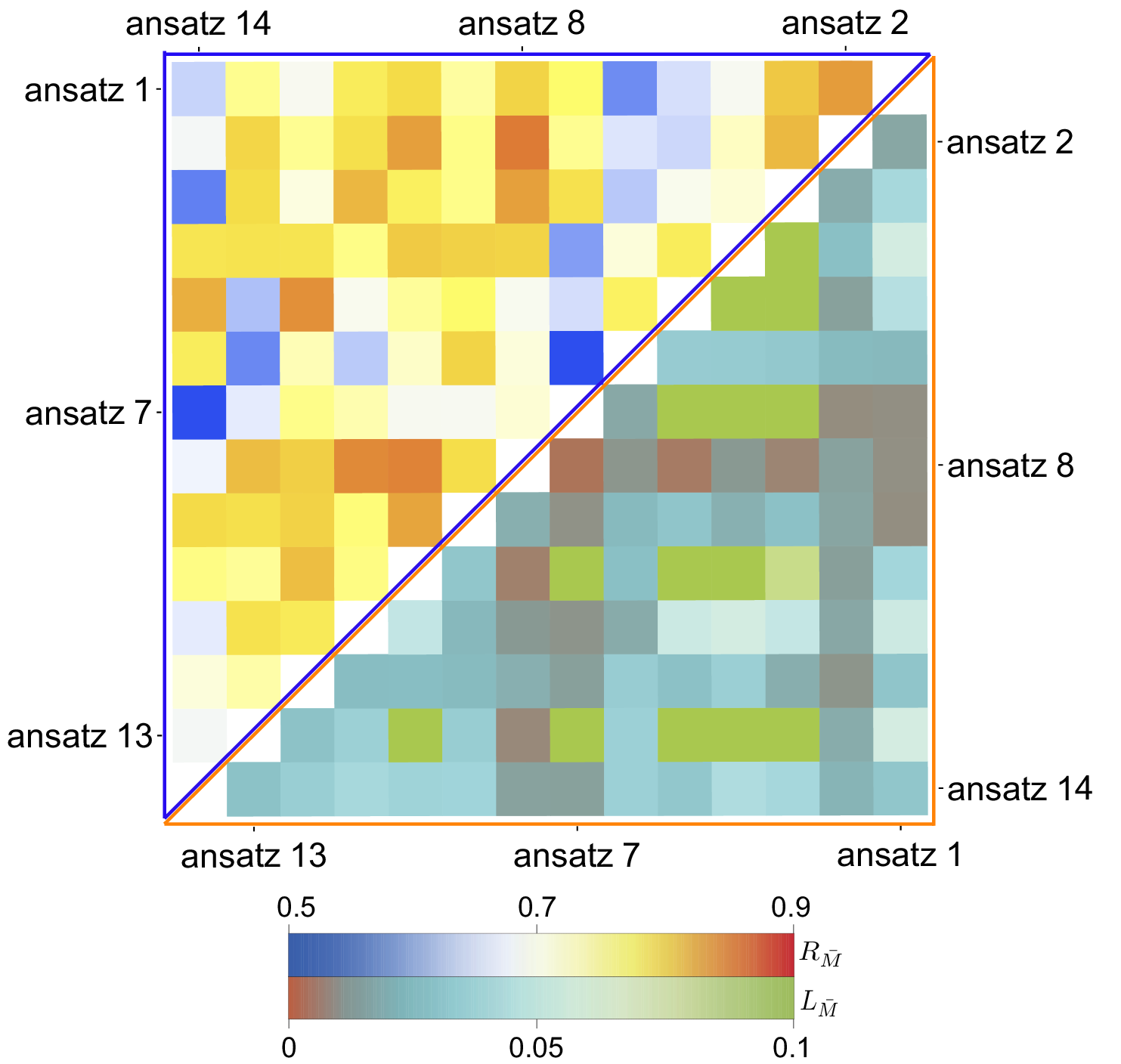}
\caption{(Color online) Plot of numerical results demonstrating the
superiority of LCA over SA. In the blue triangle, the colored block
corresponding to ansatz $\{i,j\}$ represents $R_{\{i,j\}}$. The magnitude of
$R_{\bar{M}}$ can be seen in the first color bar and indicates that LCA $%
\{i,j\}$ has higher expressibility than the highest one among ansatz $i$ and
$j$ by a percentage of $R_{\bar{M}}$. Similarly, in the brown triangle, the
colored block corresponding to ansatz $\{i,j\}$ represents $L_{\{i,j\}}.$
The magnitude of $R_{\bar{M}}$ can be seen in the second color bar and
indicates that the ground state energy estimated using LCA $\{i,j\}$ is more
accurate than the most accurate using ansatz $i$ or $j$ by a percentage of $%
L_{\bar{M}}$.}
\label{Improvementof2Ansatzes}
\end{figure}

This subsection presents a detailed comparison between LC$2$A (i.e., $M=2$)
and SA based on the numerical results of expressibility and ground state
energy estimation.

We numerically calculate the KL divergence of LC$2$A and SA to compare their
expressibilities. The numerical\ simulation involves $14$ common HEAs, with
LC$2$A representing the linear combination of every two HEAs. The
expressibility of the $i$th SA is denoted by $D_{KL}^{i}$, while that of LC$M
$A is denoted by $D_{KL}^{\bar{M}}$. The improvement in expressibility is
thus defined by
\begin{equation}
R_{\bar{M}}=\frac{D_{KL}^{\bar{M}}-\text{Min}%
[\{D_{KL}^{i},D_{KL}^{j},D_{KL}^{k},...\}]}{\text{Min}%
[\{D_{KL}^{i},D_{KL}^{j},D_{KL}^{k},...\}]}.  \label{RM}
\end{equation}%
The term Min$[{D_{KL}^{i},D_{KL}^{j},D_{KL}^{k},...}]$ denotes the maximum
expressibility among the ansatzes combined. Eq.~\eqref{Expr} demonstrates
that the expressibility of the LC$M$A for the ansatz set ${i,j,k...}$ is
improved by at least a degree of $R_{\bar{M}}$ over the SA expressibility.
For $M=2$, the numerical values of $R_{\bar{2}}$ are depicted by the blue
triangles in Fig. \ref{Improvementof2Ansatzes}. The color block for {ansatz $%
i$, ansatz $j$} signifies $R_{{i,j}}$, indicating the improvement in
expressibility of the ansatz set ${i,j}$ over the highest SA expressibility
between the $i$th and $j$th ansatzes. The magnitude of $R_{\bar{2}}$ is
indicated by the color bar.

For instance, $R_{\bar{M}}$ in the first column, where $\bar{M}=\{i,14\}$
and $i=1,...13,$ means the expressibility of ansatz $14$ has a $R_{\bar{M}
} \times 100\% $ improvement when ansatz $14$ is linearly combined with other ansatzes.
Among the $13$ ansatzes, ansatzes $1,2,3,7,8,11$, $13$ exhibit lower
improvements in expressibility than others. This can be explained by the
fact that ansatz $14$ has relatively high expressibility (see Fig.~\ref{DKLVaryingwithQandL}(a)), while the above-mentioned seven ansatzes have
lower expressibility ($D_{KL}>0.2$) than the remaining six ansatzes ($%
D_{KL}<0.2$) and ansatz $14$ ($D_{KL}<0.1$). The stronger the expressibility
of the ansatz involved in the linear combination, the greater the
improvement in the expressibility of ansatz $14$. This pattern holds true
for $R_{\bar{M}}$ in other columns. These numerical results suggest that the
expressibility of LC$2$A surpasses that of SA by $50\%$ to $90\%$,
predominantly around $80\%$.

To demonstrate LC$2$A's superiority, we compare its calculated ground state
energy with that of simulated SA using the transverse-field XY model as a
benchmark. The Hamiltonian for this model is defined as
\begin{equation}
H=\sum_{i=1}^{N}(J_{xx}X_{i}X_{i+1}+J_{yy}Y_{i}Y_{i+1})+J_{x}%
\sum_{i=1}^{N}X_{i}+J_{z}\sum_{i=1}^{N}Z_{i}
\end{equation}%
and $X_{N+1}=X_{1},Y_{N+1}=Y_{1}$. {Such a Hamiltonian does not have an eigenstate that is easy to prepare, so the previous method of measuring amplitude that does not require auxiliary qubits cannot be used~\cite{lu2021algorithms}}.

We define
\begin{equation}
L_{\bar{M}}=\frac{E_{KL}^{\bar{M}}-\text{Min}%
[\{E_{KL}^{i},E_{KL}^{j},E_{KL}^{k},...\}]}{\text{Min}%
[\{E_{KL}^{i},E_{KL}^{j},E_{KL}^{k},...\}]},
\end{equation}%
where $E_{KL}^{i}$ ($E_{KL}^{\bar{M}}$) denotes the estimated ground state energy of Hamiltonian $H$ obtained by the Variational Quantum Eigensolver
(VQE) with the $i$-th ansatz (the linear combination of ansatzes set $\bar{M}
=\{i,j,k,\ldots \}$). The notation Min$[{E_{KL}^{i},E_{KL}^{j},E_{KL}^{k},
\ldots }]$ signifies the closest approximation to the ground state energy
within the set ${E_{KL}^{i},E_{KL}^{j},E_{KL}^{k},\ldots }$. The numerical
simulation for $L_{\bar{M}}$, as illustrated by the brown triangle in Fig. %
\ref{histogram} for $M=2$, indicates that the ground state energy estimation
from LC$2$A are more precise than those from SA by $0.1\%$ to $10\%$,
predominantly around $5\%$. This suggests that enhancements in the
expressibility of ansatz sets corresponds to increased accuracy in ground
state energy estimation.

\subsection{The improvement of the expressibility for multi-ansatzs}

\label{The improvement of the expressibility for multi-ansatzs}
\begin{figure}[t]
%\centering
\includegraphics[bb=0 0 725 479, width=8 cm,clip]{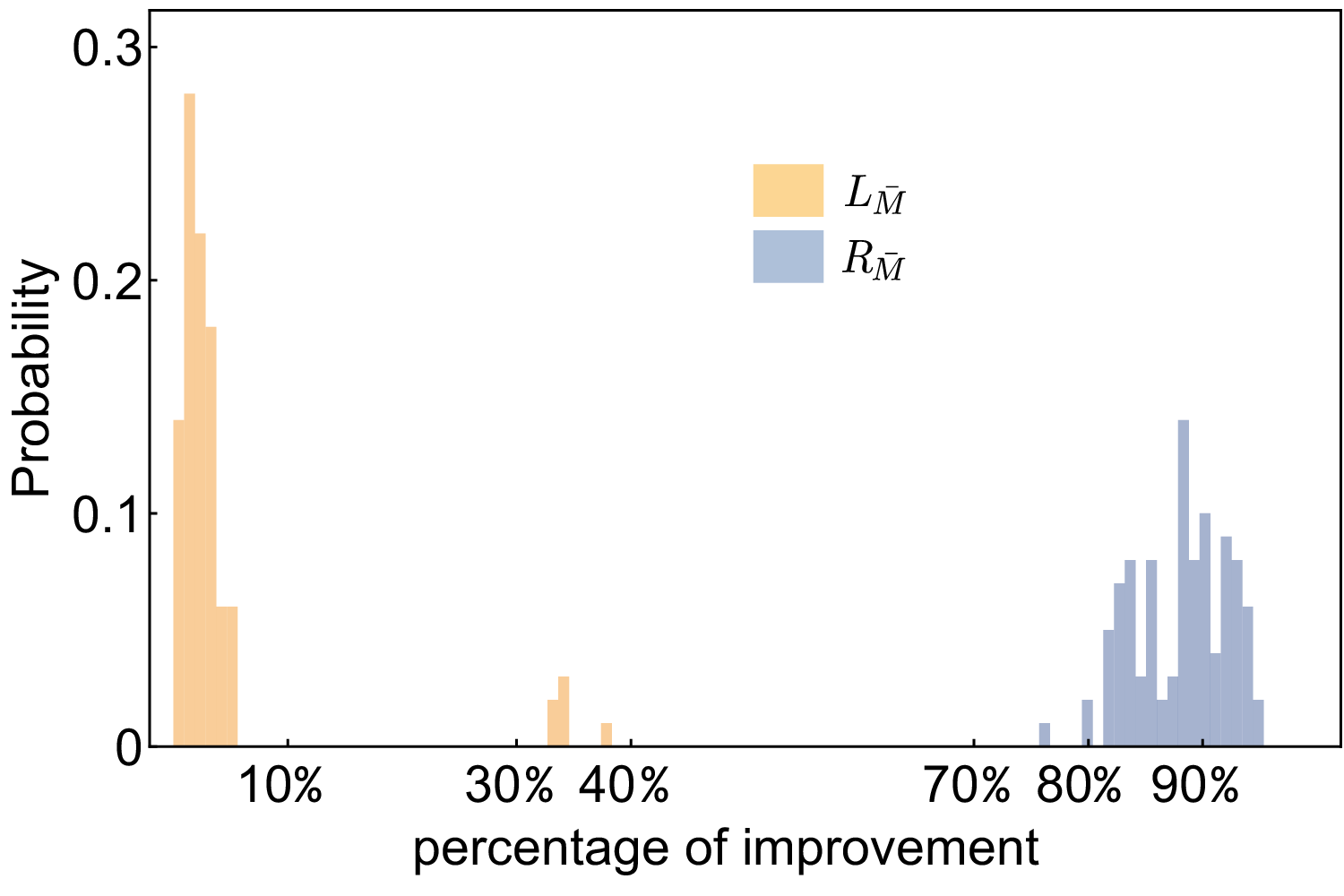}
\caption{Histogram showing the percentages linked to enhancements in
expressibility and the estimated ground state energy. Orange and blue
rectangles indicate improvements in expressibility and estimated ground
state energy, respectively, achieved by the three-ansatz LCA over the {
\color{blue}SA}, quantified by percentages $R_{\bar{M}}$ for expressibility
and $L_{\bar{M}}$ for ground state energy.}
\label{Improvementof3Ansatzes}
\end{figure}
\begin{figure*}[t]
\includegraphics[bb=0 0 1728 1649, width=15 cm, clip]{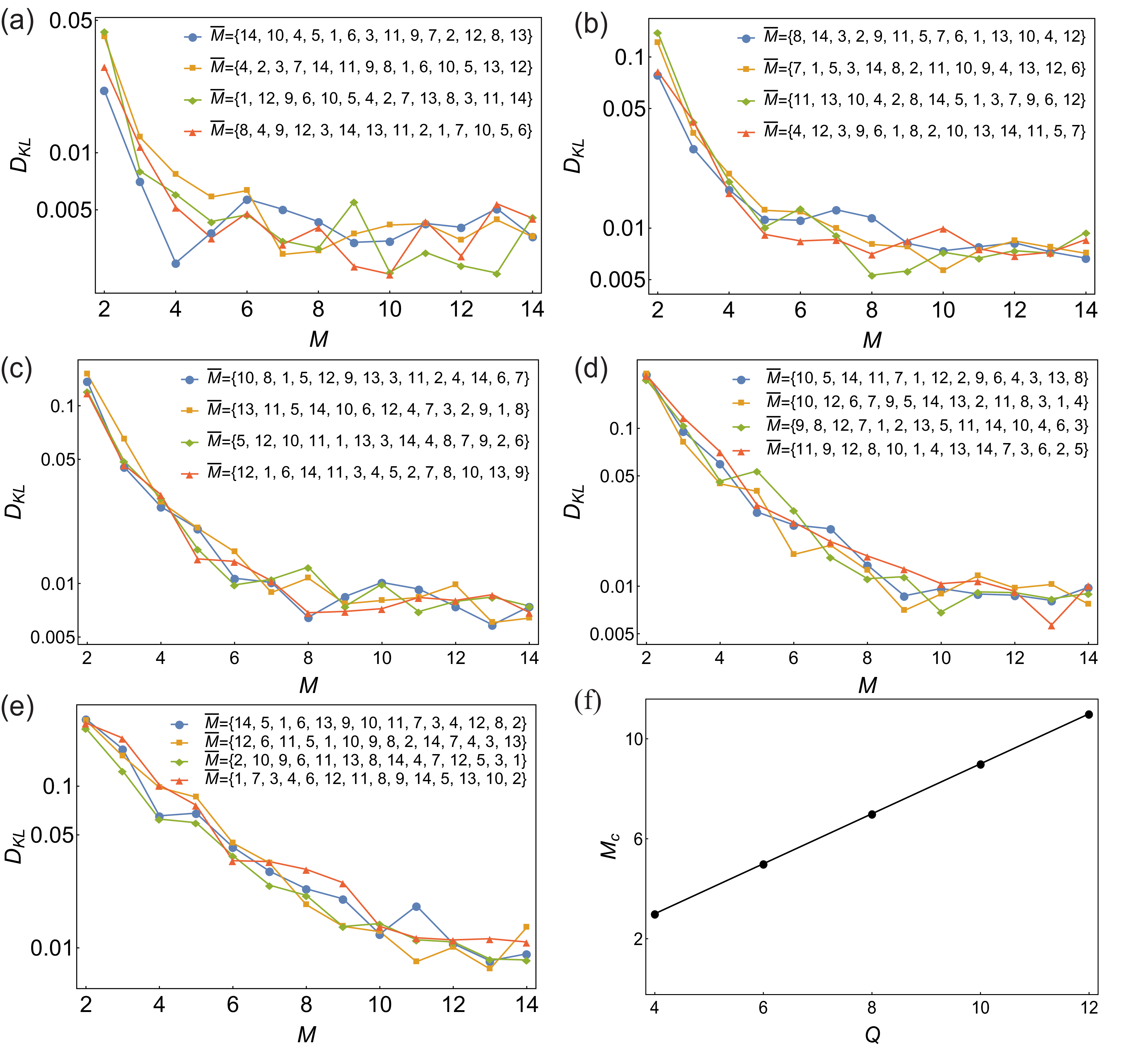}
\caption{Expressibility changes with the number of ansatzes under different
numbers of qubits. (a), (b), (c), (d), and (e) represent the numerical
results for systems with qubit numbers of 4, 6, 8, 10, and 12 respectively.
The horizontal axis is the number of ansatzes, and the vertical axis is
expressibility. Different colors represent different LCAs. (f) represents
the relation between $M_{c}$ and the number of qubits.}
\label{DKLVaryingwithQandAnsatzNumber}
\end{figure*}

This subsection evaluates the expressibility enhancement of the LCMA for $%
M\geqslant 3$. For $M=3$ and $N=4$, our numerical simulations, depicted in
Fig. \ref{Improvementof2Ansatzes}, were performed on $100$ sets of three
ansatzes randomly selected from $14$ HEAs. The improvements in
expressibility and estimated ground state energy for each set are
represented as orange and blue rectangles, respectively, in Fig. \ref%
{Improvementof2Ansatzes}. The expressibility data, with $R_{\bar{M}}$
averaging around $90\%$, demonstrates that LC$3$A's expressibility surpasses
that of SA by over $90\%$ and also indicates superior expressibility
compared to LC$2 $A. Meanwhile, the $L_{\bar{M}}$ data, averaging around $%
3\% $, suggests a modest expressibility enhancement over SA by $3\%$.
Notably, a few $L_{\bar{M}}$ values exceed $30\%$, attributed to the
relatively high ground state energy estimation by SA. Contrarily, the
improvements on ground state energy estimation offered by LC$3$A are not
markedly superior to those by LC2A, given the latter's already high
accuracy. Increasing $M$ does not lead to a continuous increase in the
expressibility of LC$M$A. $D_{\mathrm{KL}}$ of LC$M$A stabilizes when $M$
increases to the critical value $M_{c}$ such that the expressibilitiy of LC$%
M $A becomes saturated, that is, continuing to increase $M$ will not
significantly improve expressibility. The value of $M_{c}$ depends on the
system size as shown in Fig.~\ref{DKLVaryingwithQandAnsatzNumber}, where we
calculate how the expressibility of systems with qubit numbers of $4$, $6$, $%
8$, $10$, and $12$ changes with the number of ansatz. For each fixed-size
system, we performed four experiments, marked by different colors in Fig.~%
\ref{DKLVaryingwithQandAnsatzNumber}(a)-(e). We randomly selected and added
new ansatz to the LCA for each experiment until $14$ different ansatz were
selected. Fig. \ref{DKLVaryingwithQandAnsatzNumber}(f) indicates that $M_{c}$
increases almost linearly with the number of qubits.

\section{The experimental scheme of LCA}
\label{implement}
\subsection{The implementation of LCA}

\begin{figure}[t]
\centering
\includegraphics[width = 1\linewidth]{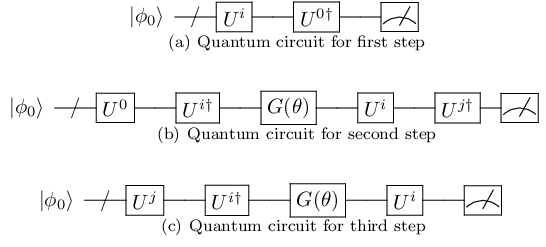}
% \subfigure[Quantum circuit for first step]  { \Qcircuit @C = 1em @R = 2em{ \lstick{\ket{\phi_{0}}} &{/}\qw &\gate{U^{i}} &\qw & \gate{U^{0\dagger}} & \qw&\meter
% } \label{circuit: overlap} }
% \par
% \subfigure[Quantum circuit for second step]{

% \Qcircuit @C = 1em @R = 2em{
% \lstick{\ket{\phi_{0}}} &{/}\qw&\gate{U^0}&\qw &\gate{U^{i\dagger}} &\qw & \gate{G(\theta)}&\qw&\gate{U^{i}} & \qw&\gate{U^{j\dagger}}&\meter
% }
% \label{circuit: cross rho}
% }
% \subfigure[Quantum circuit for third step]{

% \Qcircuit @C = 1em @R = 2em{
% \lstick{\ket{\phi_{0}}} &{/}\qw &\gate{U^j}&\qw &\gate{U^{i\dagger}} &\qw & \gate{G(\theta)}&\qw&\gate{U^{i}} & \qw&\meter
% }
% \label{fig: cross H}

% } 
\caption{Quantum circuits for LCA. (a) depicts the circuit for measuring the
overlap $\langle \protect\psi_0|\protect\psi_i\rangle$, using a final
projection measurement. Similarly, (b) demonstrates the circuit for
measuring the overlap $\langle\protect\psi_i|\protect\psi_j\rangle$, again
with a final projection measurement. Finally, (c) illustrates the circuit
for measuring $\langle\protect\psi_i|H|\protect\psi_j\rangle$, where the
final measurement employs Pauli operators $P_s$.}
\label{fig:original}
\end{figure}

To update the parameters, we must compute the mean value of the Hamiltonian:
\begin{equation}
\langle H\rangle_{\boldsymbol{\tilde{\theta}}} = \langle\psi_{\boldsymbol{%
\tilde{\theta}}}|H|\psi_{\boldsymbol{\tilde{\theta}}}\rangle = \frac{1}{%
\Omega^2} \sum_{i,j=0}^{M-1} c_i^{\dagger} c_j\langle\phi_0|U^{i\dagger}H
U^{j}|\phi_0\rangle,
\end{equation}
where $\Omega$ is the normalization constant
\begin{equation}
\Omega^2 = \sum_{i,j}^M c_i^{\dagger} c_j\langle\phi_0|U^{i\dagger}
U^{j}|\phi_0\rangle.
\end{equation}

The most common method for calculating the cross terms $\langle\phi_0|U^{i
\dagger}H U^{j}|\phi_0\rangle$ and $\langle\phi_0|U^{i\dagger}
U^{j}|\phi_0\rangle$ where $i\neq j $ is the Hadamard test(HT). HT requires
the implementation of control-$U^i$ gate, gates, which typically involve
numerous two-qubit gates. In real devices, these contribute significantly to
error accumulation. To mitigate this overhead, we propose PCM specifically
designed for LCA. PCM eliminates the need for auxiliary qubits and minimizes
two-qubit gate usage. We subsequently detail efficient methods for measuring
both $\langle\psi_i|\psi_j\rangle$ and $\langle\psi_i|H|\psi_j \rangle$.

\begin{enumerate}
\item The algorithm begins by calculating $\mathrm{Tr}(\rho_0\rho_i)$ for $i
= 1,2,\dots,M-1$, where $\rho_i = |\psi_i\rangle\langle\psi_i| =
U^{i\dagger}|\phi_0\rangle\langle\phi_0|U^i$. Although $\mathrm{Tr}(\rho_0
\rho_i) = |\langle\psi_0|\psi_i\rangle|^2$, we initially assume $\langle
\psi_0|\psi_i\rangle = \sqrt{\mathrm{Tr}(\rho_0\rho_i)}$ in this step. In
fact, this equation does not necessarily hold, because there may be a
relative phase between the two states. This is valid despite potential
relative phases between the states, as the next section will justify their
irrelevance in this context. The specific implementation circuit diagram is
shown in Fig.~\ref{fig:original}(a).

\item The next step is to measure $\mathrm{Tr}(\rho_0\rho_i\rho_j)$ for $i,j
= 1,2,\dots,M-1$ and $i\neq j$ using the circuit in Fig.~\ref{fig:original}(b). To obtain $\mathrm{Tr}(\rho _{0}\rho _{i}\rho _{i})$ we implement the
circuit consisting of \{$U_{0},(U_{i})^{\dagger },I-x\left\vert
0\right\rangle \left\langle 0\right\vert ,U_{i},(U_{j})^{\dagger }$\}, where
$x = 1-e^{i\theta}$ and $I - x|0\rangle\langle 0| = e^{i\theta
|0\rangle\langle 0|}$. The unitary operator applied to $\rho$ is actually $%
U_i(I-x|0\rangle\langle 0|)U_i^{\dagger} = I - x|\psi_i\rangle\langle\psi_i|
= I-x\rho_i$. What the circuit outputs are
\begin{eqnarray}
&\mathrm{Tr}((I - x\rho_i) \rho_0 (I - x^{*}\rho_i)\rho_j) \\
=& \mathrm{Tr} (\rho_0\rho_j-x\rho_i\rho_0\rho_j - x^{*}\rho_0\rho_i\rho_j +
|x|^2\rho_i\rho_0\rho_i\rho_j ).
\end{eqnarray}
Note that $\mathrm{Tr}(\rho_0\rho_j)$ was already obtained and $\mathrm{Tr}%
(\rho_i\rho_0\rho_i\rho_j) = \mathrm{Tr}(\rho_i\rho_0)\mathrm{Tr}
(\rho_i\rho_j)$ so they can be calculated from Fig.~\ref{fig:original}(b).
To extract $\mathrm{Tr}(\rho_0\rho_i \rho_j) $ from the measurements, we use
two different $x$ values, treating treat $\mathrm{Tr}(\rho_0\rho_i\rho_j)$
and $\mathrm{Tr}(\rho_i\rho_0\rho_j)$ as unknowns and solving the linear
system. In our numerical simulations, we set $x=2$ and $x=1-i$ respectively.
The resulting system of equations satisfy the following form:
\begin{equation*}
\left\{
\begin{array}{c}
\mathrm{Tr}(2\rho _{0}\rho _{i}\rho _{j} + 2\rho_{i}\rho _{0}\rho_{j})=R_{1}
\\
\mathrm{Tr}((1+i)\rho _{0}\rho _{i}\rho _{j}+(1-i)\rho _{i}\rho _{0}\rho
_{j})=R_{2}%
\end{array}
\right. ,
\end{equation*}
Where $R_1$ and $R_2$ are derived from the measurements. This system has a
unique solution. With the extracted $\mathrm{Tr}(\rho_0\rho_i\rho_j)$, we
can calculate $\langle\psi_i|\psi_j\rangle$ form the relation
\begin{eqnarray*}
\langle\psi_i|\psi_j\rangle = &\frac{\langle\psi_0|\psi_i\rangle\langle
\psi_i|\psi_j\rangle\langle\psi_j|\psi_0\rangle}{\langle\psi_0|\psi_i\rangle
\langle\psi_j|\psi_0\rangle} \\
=& \frac{\mathrm{Tr}(\rho_0\rho_i\rho_j)}{\sqrt{\rho_0\rho_i}\sqrt{
\rho_0\rho_j}}.
\end{eqnarray*}

\item Given the Hamiltonian $H = \sum_{s}a_{s}P_s$, where $P_s$ are
Hermitian operators, we can estimate $\langle\psi_i|P_s|\psi_j\rangle$ using
the circuit in Fig.~\ref{fig:original}(c).This is similar to the previous step
but with the final measurement replaced by $P_s$. The measurement results
yield:
\begin{eqnarray*}
&\mathrm{Tr}((I-x\rho_j)\rho_i(I-x^{*}\rho_j)P_s) \\
=& \mathrm{Tr}(\rho_i P_s -x\rho_j\rho_i P_s - x^{*}\rho_i\rho_j P_s
+|x|^2\rho_j\rho_i\rho_j P_s)
\end{eqnarray*}
Both $\mathrm{Tr}(\rho_i P_s)$ and $\mathrm{Tr}(\rho_j\rho_i\rho_j P_s) =
\mathrm{Tr}(\rho_j\rho_i)\mathrm{Tr}(\rho_j P_s)$ can be obtained directly
from other quantum circuits. Similar to the previous approach, we use
different $x$ values and solve the system of equations:
\begin{equation*}
\left\{
\begin{array}{c}
\mathrm{Tr}(2\rho _{j}\rho _{i}P_s+ 2\rho_{i}\rho _{j}P_s)=R_{1}^{\prime} \\
\mathrm{Tr}((1+i)\rho _{j}\rho _{i}P_s+(1-i)\rho _{i}\rho
_{j}P_s)=R_{2}^{\prime}%
\end{array}%
, \right.
\end{equation*}
Finally we calculate $\langle\psi_i|P_s|\psi_j\rangle$ according to the
relation
\begin{eqnarray*}
\langle\psi_i|P_s|\psi_j\rangle =& \frac{\langle\psi_j|\psi_i\rangle\langle
\psi_i|P_s|\psi_j\rangle}{\langle\psi_j|\psi_i\rangle} \\
=& \frac{\rho_j\rho_i P_s}{\langle\psi_j|\psi_i\rangle},
\end{eqnarray*}
where $\langle\psi_j|\psi_i\rangle$ is obtainable from the previous step.
\end{enumerate}

\subsection{Unknown phase}

In the previous section, we assumed phaseless $\langle\psi_0|\psi_i\rangle$.
Here, we justify this assumption. Suppose a method like HT accurately
determines the phase of $\langle\psi_0|\psi_i\rangle$. Using the variational
method, we minimize the energy to obtain solutions $\{c_{i},\boldsymbol{%
\theta}^{i}\}$. Suppose the phase of $\langle\psi_0|\psi_j\rangle$ is $%
e^{i\alpha_i}$ . Ignoring the phase during calculations effectively applies
a transformation :
\begin{eqnarray}
|\psi_i\rangle \rightarrow |\psi_i^{\prime}\rangle =
e^{-i\alpha_i}|\psi_i\rangle
\end{eqnarray}
The calculated average energy then becomes:
\begin{eqnarray}
\langle H\rangle_{\tilde{\boldsymbol{\theta}}} = \langle\psi_{\tilde{%
\boldsymbol{\theta}}}^{\prime}|H|\psi_{\tilde{\boldsymbol{\theta}}%
}^{\prime}\rangle = \frac{1}{\Omega^2} \sum_{i,j=0}^{M-1} c_i^{\dagger}
c_j\langle\psi_i^{\prime}|H|\psi_j^{\prime}\rangle.
\end{eqnarray}
Thus, there still exists a set of solutions $\{c_{i}e^{i\alpha_i},%
\boldsymbol{\theta}^i \}$ that yields the same minimum energy value as the
HT method. In this sense, we can consider that the assumption of no phase in
$\langle\psi_0|\psi_j\rangle$ has no impact on the algorithm.

\subsection{Two qubits gate counting}

A key advantage of PCM over HT lies in the markedly reduced need for
two-qubit gates. In HT, each CNOT gate in the ansatz transforms into a
Toffoli gate, which then decomposes into five two-qubit gates. Conversely,
in PCM, CNOT gates in the ansatz are preserved, though the number of
two-qubit gates for implementing the $G(\theta)$ gate escalates with the bit
count. This distinction is critical for NISQ-era hardware.

\begin{figure}[tbp]
\includegraphics[bb=0 0 749 977, width=8 cm,clip]{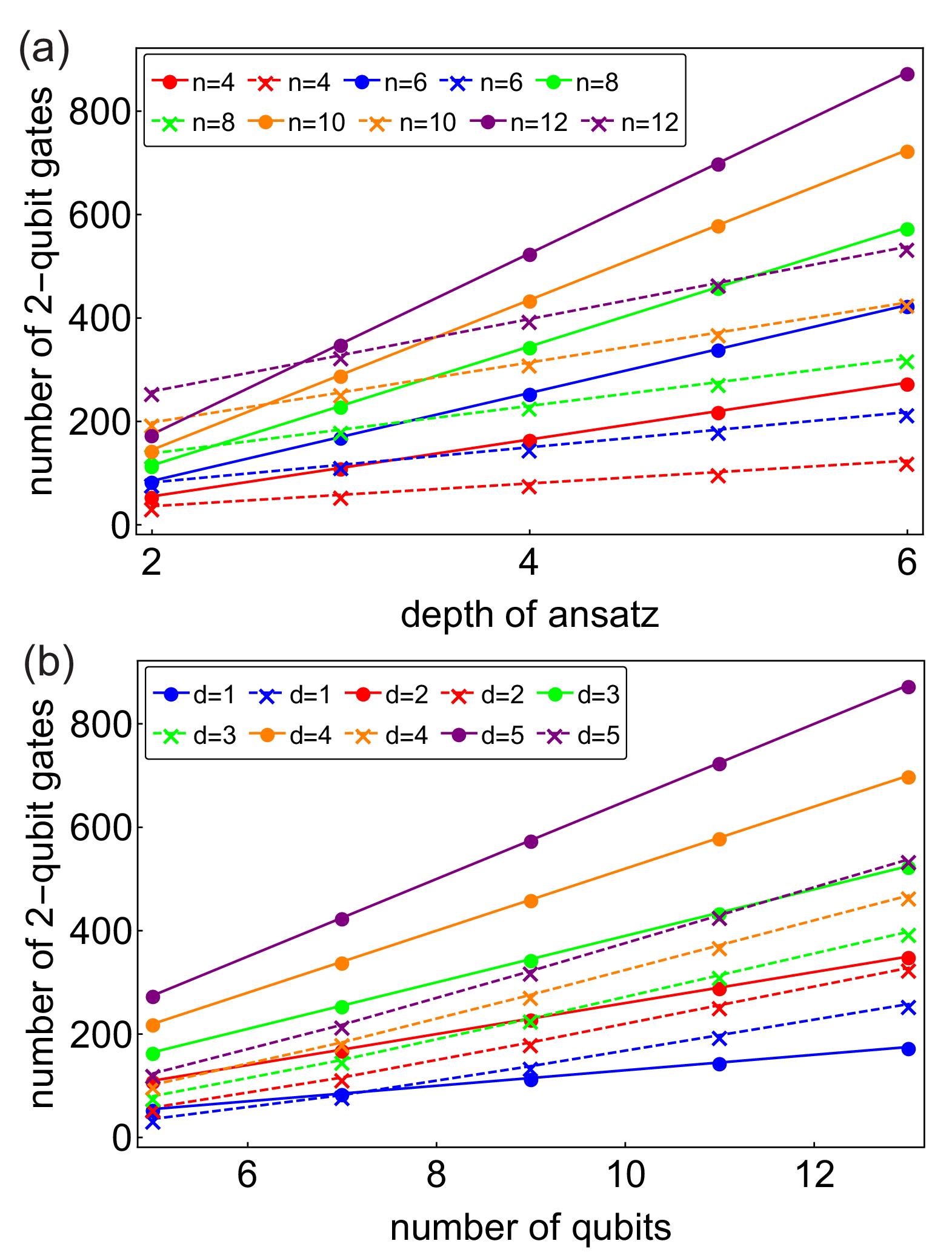}
\caption{Comparison of PCM and HT. The solid lines represent data for the HT
method, while the dashed lines represent data for the PCM method. (a) The
horizontal axis denotes the ansatz depth, defined as the number of
repetitions of circuits 2 and 15 within the overall circuit. The vertical
axis represents the two-qubit gate count. The number of qubits, $n$, is
indicated by lines of different colors. (b) The number of qubits is shown on
the horizontal axis, and the vertical axis displays the two-qubit gate
count. The depth of ansatz, $d$, is represented by lines of different
colors. }
\label{2qGates}
\end{figure}

We take circuit-2 and circuit-15 as an example of LCA and compare the number
of two-qubit gates required by the two methods, as shown in Fig.~\ref%
{2qGates}(a) and Fig.~\ref{2qGates}(b). An $n$-qubit $G(\theta)$ gate is
implemented by a series of single-qubit gates and an $n$-qubit Toffoli gate.
This Toffoli gate is implemented using the method in Ref.~\cite{5954249}
(Note that some other, better implementations of Toffoli gates~\cite%
{nie2024quantum} would bring more advantages to PCM). When the ansatz is
shallow and the number of qubits is large, PCM has no advantage over HT.
However, this scenario is often less relevant in practical applications. To
achieve an accurate approximation of the desired state, the ansatz depth
typically scales with the number of qubits. As shown in Fig.~\ref{2qGates}%
(a), for a fixed number of qubits, the advantage of PCM over HT becomes
increasingly pronounced with increasing ansatz depth. Furthermore, Fig.~\ref%
{2qGates}(b) demonstrates that when the ansatz depth exceeds a critical
threshold($d\geq 3$), PCM exhibits a significant reduction in two-qubit
gates required compared to HT as the number of qubits increases. Although
this threshold will change slightly depending on the choice of ansatz,
within the regime where the quantum advantage is achievable(number of qubits
$> 50$), the performance of PCM is anticipated to significantly surpass that
of HT.

\subsection{Parameter shift rules}

A straightforward approach to the optimization of variational algorithms is
gradient descent, which requires the estimation of gradients. The gradients
for $\boldsymbol{c}$ are relatively simple:
\begin{equation}
\begin{aligned} \frac{\partial \langle H\rangle}{\partial c_i} =
\frac{1}{|\langle\psi_{\Tilde{\boldsymbol{\theta}}}|\psi_{\Tilde{%
\boldsymbol{\theta}}}\rangle|^2}\left(
\langle\psi_{\tilde{\boldsymbol{\theta}}}|H|\psi_{i}\rangle + h.c. \right).
\end{aligned}
\end{equation}%
The gradients for $\boldsymbol{\theta }$ are relatively complex:
\begin{equation}
\begin{aligned} \frac{\partial \langle H\rangle}{\partial \theta_{i,l}} =
\frac{1}{|\langle\psi_{\Tilde{\boldsymbol{\theta}}}|\psi_{\Tilde{%
\boldsymbol{\theta}}}\rangle|^2}&
\left(\frac{\partial\langle\psi_{\Tilde{\boldsymbol{\theta}}}|H|\psi_{%
\Tilde{\boldsymbol{ \theta}}}\rangle}{\partial
\theta_{i,l}}\langle\psi_{\Tilde{\boldsymbol{
\theta}}}|\psi_{\Tilde{\boldsymbol{ \theta}}}\rangle\right.\\ &\left.-
\langle\psi_{\Tilde{\boldsymbol{ \theta}}}|H|\psi_{\Tilde{\boldsymbol{
\theta}}}\rangle \frac{\partial\langle\psi_{\Tilde{\boldsymbol{
\theta}}}|\psi_{\Tilde{\boldsymbol{ \theta}}}\rangle}{\partial \theta_{i,l}}
\right), \label{eq: derivation} \end{aligned}
\end{equation}%
where
\begin{equation}
\begin{aligned} \frac{\partial\langle\psi_{\Tilde{\boldsymbol{
\theta}}}|\psi_{\Tilde{\boldsymbol{ \theta}}} \rangle} {\partial
\theta_{i,l}} &= \sum_{j}c_i^{*} c_j \frac{\partial
\langle\psi_{i}|}{\partial \theta_{i,l}}|\psi_{j}\rangle + h.c.\\ & =
\sum_{j}c_i^{*} c_j \langle\phi_{0}|\frac{\partial U^{i\dagger}}{\partial
\theta_{i,l}}U^{j}|\phi_{0}\rangle + h.c., \label{eq: ddd} \end{aligned}
\end{equation}%
and
\begin{equation}
\begin{aligned} \frac{\partial\langle\psi_{\Tilde{\boldsymbol{
\theta}}}|H|\psi_{\Tilde{\boldsymbol{ \theta}}}\rangle} {\partial
\theta_{i,l}} &= \sum_{j}c_i^{*} c_j \frac{\partial
\langle\psi_{i}|H|}{\partial \theta_{i,l}}|\psi_{j}\rangle + h.c.\\ & =
\sum_{j}c_i^{*} c_j \langle\phi_{0}|\frac{\partial U^{i\dagger}}{\partial
\theta_{i,l}}HU^{j}|\phi_{0}\rangle + h.c.. \label{eq: H ddd} \end{aligned}
\end{equation}%
As mentioned in the previous subsection, when we estimate the normalization $%
\langle \psi _{\Tilde{\boldsymbol{ \theta}}}|\psi _{\Tilde{\boldsymbol{
\theta}}}\rangle $ average the Hamiltonian $\langle \psi _{\Tilde{%
\boldsymbol{ \theta}}}|H|\psi _{\Tilde{\boldsymbol{ \theta}}}\rangle $, we
will change the final state by unknown phases. When estimating Eq.~%
\eqref{eq: ddd} and Eq.~\eqref{eq: H ddd}, we cannot introduce more unknown
phases to ensure correct results. This can be achieved when the $t$-th level
of $U^{i\dagger }$ is of the form $U_{t}^{i\dagger }=e^{iO_{i,t}\theta
_{i,t}}$, where $O_{i,l}$ is a known Hermitian operator, e.g., Pauli
operator. Let $U^{i\dagger }=\prod_{t}U_{t}^{i\dagger }(\theta _{i,t})$,
then the partial differentiation ${\partial U_{i}^{\dagger }}/{\partial
\theta _{i,l}}$ can be written as
\begin{equation}
\begin{aligned} \frac{\partial U^{i\dagger}}{\partial \theta_{i,l}} =
\prod_{t=1}^l U^{i\dagger}_{t}(\theta_{i,t}) O_{i,l}\prod_{t=l+1}
U^{i\dagger}_{t}(\theta_{i,t}). \end{aligned}
\end{equation}%
We first consider the inner product in Eq.~\eqref{eq: ddd} in the case where
$i=j$ can be expressed as
\begin{equation}
\begin{aligned} \langle\phi_{0}|\frac{\partial U^{i\dagger}}{\partial
\theta_{i,l}}U^{i}|\phi_{0}\rangle &= \langle\phi_0|\prod_{t=1}^l
U^{i\dagger}_{t}(\theta_{i,t}) O_{i,l} \prod_{t=l}^1
U^{i\dagger}_{t}(\theta_{i,t}) |\phi_0\rangle\\ & = \langle
\psi_{i,l}|O_{i,l}|\psi_{i,l}\rangle, \end{aligned}
\end{equation}%
which can be estimated directly using quantum circuits. Then we can estimate
the inner product in the general case where $i\neq j$ from
\begin{equation}
\begin{aligned} \langle\phi_{0}|\frac{\partial U^{i\dagger}}{\partial
\theta_{i,l}}|\psi_{i}\rangle\langle\psi_{i}|\psi_{j}\rangle\langle\psi_{j}|
\frac{\partial U^{i}}{\partial \theta_{i,l}}|\phi_0\rangle, \end{aligned}
\end{equation}%
which can be estimated using Step 2 mentioned in the last section. We use
\begin{equation}
\begin{aligned} \langle\phi_{0}|\frac{\partial U^{i\dagger}}{\partial
\theta_{i,l}}|\psi_{j}\rangle\langle\psi_{j}|H| \frac{\partial
U^{i}}{\partial \theta_{i,l}}|\phi_0\rangle, \label{eq: cross H}
\end{aligned}
\end{equation}%
to estimate $\langle \psi _{j}|H|(\partial U^{i}/\partial \theta
_{i,l})|\phi _{0}\rangle $. Eq.~\eqref{eq: cross H} can be estimated using Step 3 mentioned in the last section.

\section{SUMMARY}

To overcome the limitations of traditional approaches, we develop a novel
method for enhancing the expressibility of variational quantum circuits for
noisy quantum computers. Our key innovation lies in combining a linear
combination of ansatzes with a specialized measurement scheme, effectively
sidestepping the detrimental effects of increased circuit depth and noisy
Hadamard tests. This translates to reduced dependence on two-qubit gates,
leading to potentially more robust quantum computations. Furthermore, the
method is readily applicable due to its efficient gradient calculation
scheme. Numerical simulations validate the effectiveness of our approach,
paving the way for practical applications on current noisy quantum hardware. %
\label{Summary}

\begin{acknowledgments}
\label{acknowledgements}
We thank Wencheng Zhao for illuminating discussions. This work is supported by the National Natural Science Foundation of China (Grant No. 12305018, 12225507, 12088101), and the Fundamental Research Funds for the Central Universities (2023MS079).
\end{acknowledgments}

\bibliographystyle{plain}
\bibliography{ref.bib}

\end{document}